\documentclass[onecolumn,preprintnumbers,amsmath,amssymb,superscriptaddress,floatfix]{revtex4}
\usepackage{amsfonts}
\usepackage{amssymb}
\usepackage{amsmath}
\usepackage{amsthm}
\usepackage{graphicx}
\usepackage[dvipsnames]{xcolor}
\usepackage{bbm}
\usepackage{latexsym}

\newcommand{\lb}{\langle} 
\newcommand{\rb}{\rangle}
\newcommand{\lp}{\left(}
\newcommand{\rp}{\right)}
\newcommand{\lpq}{\left[}
\newcommand{\rpq}{\right]}
\newcommand{\alphaN}{\alpha_N}
\newcommand{\vargauss}{\mathcal{N}(0,1)}  
\newcommand{\sumi}{\sum_{i=1}^N}
\newcommand{\sumij}{\sum_{i,j=1}^N}
\newcommand{\summu}{\sum_{\mu=1}^{p-1}}

\newcommand{\si}{\sigma_i}
\newcommand{\sj}{\sigma_j}
\newcommand{\zmu}{z_\mu}
\newcommand{\xiimu}{\xi_i^\mu}
\newcommand{\xijmu}{\xi_j^\mu}

\newcommand{\tmu}{\theta_\mu}
\newcommand{\ei}{\eta_i}
\newcommand{\MG}{\mathcal{M}}
\newcommand{\E}{\mathbb{E}}  
\newcommand{\R}{\mathbb{R}} 
\newcommand{\ZN}{Z_N(\beta)}   

\newcommand{\qb}{\bar{q}}
\newcommand{\pb}{\bar{p}}
\newcommand{\xiinu}{\tilde{\xi}_i^\nu}
\newcommand{\xijnu}{\tilde{\xi}_j^\nu}
\newcommand{\xnu}{x_\nu}
\newcommand{\sumnu}{\sum_{\nu=1}^{k}}
\newcommand{\summup}{\sum_{\mu=1}^{p}}
\newcommand{\xit}{\tilde{\xi}}
\newcommand{\Zt}{\tilde{Z}_N (t,x)} 
\newcommand{\Ztx}{Z_N (t,x,\psi)} 
\newcommand {\diagonale}{- O \lp \frac{\ln N}{N} \rp - \frac{\alphaN \beta}{2}}   
\newcommand{\mnu}{m_\nu}

\newtheorem{proposition}{Proposition}

\newtheorem{theorem}{Theorem}
\newtheorem{corollario}{Corollary}
\newtheorem{remark}{Remark}
\newtheorem{conj}{Conjecture}

\newtheorem{defi}{Definition}

\def\be{\begin{equation}}
\def\ee{\end{equation}}
\def\bea{\begin{eqnarray}}
\def\eea{\end{eqnarray}}

\def\s{\sigma}

\def\e{\varepsilon}
\def\epsilon{\e}

\def\b{\beta}

\def\R{\mathbb{R}}
\def\E{\mathbb{E}}

\newcommand{\nocontentsline}[3]{}
\newcommand{\tocless}[2]{\bgroup\let\addcontentsline=\nocontentsline#1{#2}\egroup}

\DeclareMathSymbol{\leqslant}{\mathalpha}{AMSa}{"36} 
\DeclareMathSymbol{\geqslant}{\mathalpha}{AMSa}{"3E} 
\DeclareMathSymbol{\eset}{\mathalpha}{AMSb}{"3F}     
\renewcommand{\leq}{\;\leqslant\;}                   
\renewcommand{\geq}{\;\geqslant\;}                   

\begin{document}

\title{Neural Networks retrieving Boolean \\ patterns in a sea of Gaussian ones}

\author{Elena Agliari}
\affiliation{Dipartimento di Matematica, Sapienza Universit\`a di Roma, Italy}
\affiliation{Istituto Nazionale di Alta Matematica (GNFM-INdAM), Roma, Italy}
\author{Adriano Barra}
\affiliation{Dipartimento di Matematica e Fisica ``Ennio De Giorgi'', Universit\`a del Salento, Italy}
\affiliation{Istituto Nazionale di Alta Matematica (GNFM-INdAM), Roma, Italy}
\author{Chiara Longo}
\affiliation{Dipartimento di Matematica, Sapienza Universit\`a di Roma, Italy}
\author{Daniele Tantari}
\affiliation{Scuola Normale Superiore, Centro Ennio De Giorgi, Italy.}
\affiliation{Istituto Nazionale di Alta Matematica (GNFM-INdAM), Roma, Italy}

\begin{abstract}
Restricted Boltzmann Machines are key tools in Machine Learning and are described by the energy function of  bipartite spin-glasses. From a statistical mechanical perspective, they share the same Gibbs measure of Hopfield networks for associative memory. In this equivalence, weights  in the former play as patterns in the latter. As Boltzmann machines usually require real weights to be trained with gradient descent like methods, while Hopfield networks typically store binary patterns to be able to retrieve, the investigation of a \textit{mixed} Hebbian network, equipped with both real (e.g., Gaussian) and discrete (e.g., Boolean) patterns naturally arises.
\newline
We prove  that, in the challenging regime of a high storage of real patterns, where retrieval is forbidden,  an extra load of boolean patterns can still be retrieved, as long as the ratio among the overall load  and the network size does not exceed a critical threshold, that turns out to be the same of the standard Amit-Gutfreund-Sompolinsky theory. Assuming replica symmetry, we study the case of a low load of boolean patterns  combining the stochastic stability and Hamilton-Jacobi interpolating techniques. The result can be extended to the high load by a non rigorous but standard replica computation argument.
\end{abstract}

\maketitle

\section{Introduction}

In recent years we have witnessed a formidably fast development of research in Artificial Intelligence. Neural networks are playing an important role in this trend, mainly due to the ability of the so-called deep networks to solve difficult problems, upon a proper training. Such problems are broadly ranged in sciences (from Particle Physics \cite{[1]} to Computational Biology \cite{[2]}), not to mention the applied world of technology, where their usage has become pervasive. Nevertheless, as admitted in \cite{[3]}, despite its remarkable successes, nobody yet understands exhaustively how the whole scaffold works, while there is wide agreement that achieving a full understanding of Deep Learning is an urgent priority.
\newline
The pivotal constituent of Deep Learning machinery is the Restricted Boltzmann Machine (RBM) \cite{hinton1,hinton2,RBM1,RBM2}. This is a network of units with a bipartite structure, the two parties being referred to as {\em visible layer} and {\em hidden layer}; units belonging to different layers are connected by links endowed with {\em weights} while nodes belonging to the same layer are not connected  (see Fig.$1$ left panel).  In the jargon of statistical physicists, RBMs have the same energy of a bipartite spin-glass \cite{bg,Bip,zigg,pizzo,auffinger}.

By marginalization over the hidden layer, RBMs have also been shown to share the same phase diagram of an Hopfield network \cite{BBCS,prlnoi1,mezard,multi,remi}, whose units, corresponding to those of the visible layer  (see Fig.$1$ right panel), are connected via an Hebbian coupling \cite{Hebb}, with number of patterns corresponding to the amount of hidden units.  The Hopfield network is able to spontaneously retrieve such patterns, and therefore to work as an associative memory\cite{ags1}, as long as the ratio between the patterns to handle and the available neurons is not too large \cite{ags2}, or, in the dual perspective of the RBMs, until the size of the hidden layer is not too large compared to the visible layer's  one. 

Crucially, the {\em weight vectors} learnt by the RBM after training play as patterns in Hopfield retrieval.
Since standard Hopfield networks are built with Boolean patterns, studies on possible generalizations are needed and begin to appear in the literature  \cite{lettera,remi}.

In the last years, an increasing number of semi-heuristic routes toward a rationale for Deep Learning have been introduced, while rigorous answers (e.g., avoiding the usage of the so called replica trick \cite{MPV,Ton,seung}) to specific questions are hardly distilled (see e.g. \cite{BGG-JSP2010,BGGT-JSM2012,S,ST,BG5,BGP3,Tala1,Tala2,lecun}). However, beyond the replica-trick, other techniques (from cavity or message passing \cite{mezard,huang,multi} to those based on interpolating structures \cite{BGGT-JSM2012,bg,Bip,prlnoi3}) to handle spin-glasses have recently appeared in the literature, hence an attempt should be made in using them to infer properties of these Restricted Boltzmann Machines also from a rigorous perspective.
\newline
Here we prove, at the replica symmetric level, that Hopfield networks endowed with patterns that are {\em mixed}, namely in part binary and in part real, are robustly capable of retrieving the digital information (i.e., the binary patterns) although ``immersed'' in the continuous  (slow) noise generate by the real patterns (i.e., the {\em sea}). In particular, in this paper, by mixing two mathematical approaches, namely stochastic stability \cite{AC-SS,BGG-JSP2010,BGGT-JSM2012,BGGTgauss} and Hamilton-Jacobi interpolation \cite{HJ-Jstat,ABDG,BDT2013,guerra-HJ,GTmult}, we are able to describe the model free energy and phase diagram for pure state retrieval.

Let us consider a system made of $N$ Ising neurons dealing with a certain number of patterns, referred to as $p$ or $k$ according to whether the number scales linearly with $N$ (i.e., $p = \alpha N$) or logarithmically with $N$ ($k = \gamma \ln N$). These two cases correspond to the so-called {\em high storage} and {\em low storage} regimes, respectively \cite{AMIT}.
As well known, in the low-storage regime the Hopfield model is able to retrieve patterns (i.e., to work as a distributed associative memory) for binary as well as real patterns \cite{lettera,lungo}, while, in the high-storage regime, only binary patterns can be retrieved because a linearly extensive (in $N$) amount of real patterns contains too much information for the $O(N^2)$ synapses to perform pattern recognition or similar tasks \cite{bov-stoc,lungo}. Indeed, in general, the high-storage case is much more tricky due to its intrinsic glassiness, whence tools from disordered statistical mechanics are in order to infer its properties \cite{AMIT,MPV}. On the contrary, standard statistical mechanical machineries are usually effective for the low-storage case \cite{Ton}.
\newline
Now, given the equivalence between RBM and Hopfield networks, a natural interest for mixed Hebbian networks (where patterns are in part analog and in part digital) arises and a first scenario we would figure out and clarify is their retrieval capabilities when they are constrained to keep an extensive amount of $p$ real patterns (hence the worst case for retrieval) but they are also over-fed by a further low-load of $k$ binary patterns.
\newline
Exploiting Guerra's interpolating schemes we prove there exists a region in the parameter space (corresponding to not-too-high values of both fast and slow noises), where  mixed Hebbian network works as a distributed associative memory and the boundaries of such a region are evidenced by a first-order phase transition.
\newline
Further, a fairly standard replica calculation, although not rigorous, suggests that this picture can be extended even to the case of an extensive load for both binary and real patterns, that is, there exists a retrieval region where pattern recognition for high-load digital information in a real sea seems possible.
\newline
Remarkably, in all these cases, the boundary for the retrieval region turns out to be always the one identified by Amit-Gutfreund-Sompolinsky in the $80$'s \cite{ags1,ags2}.

\subsection{Associative Hopfield Networks and Restricted Boltzmann Machines} \label{hnn}
\begin{figure*}[t]
\begin{center}
\includegraphics[width=.70\textwidth]{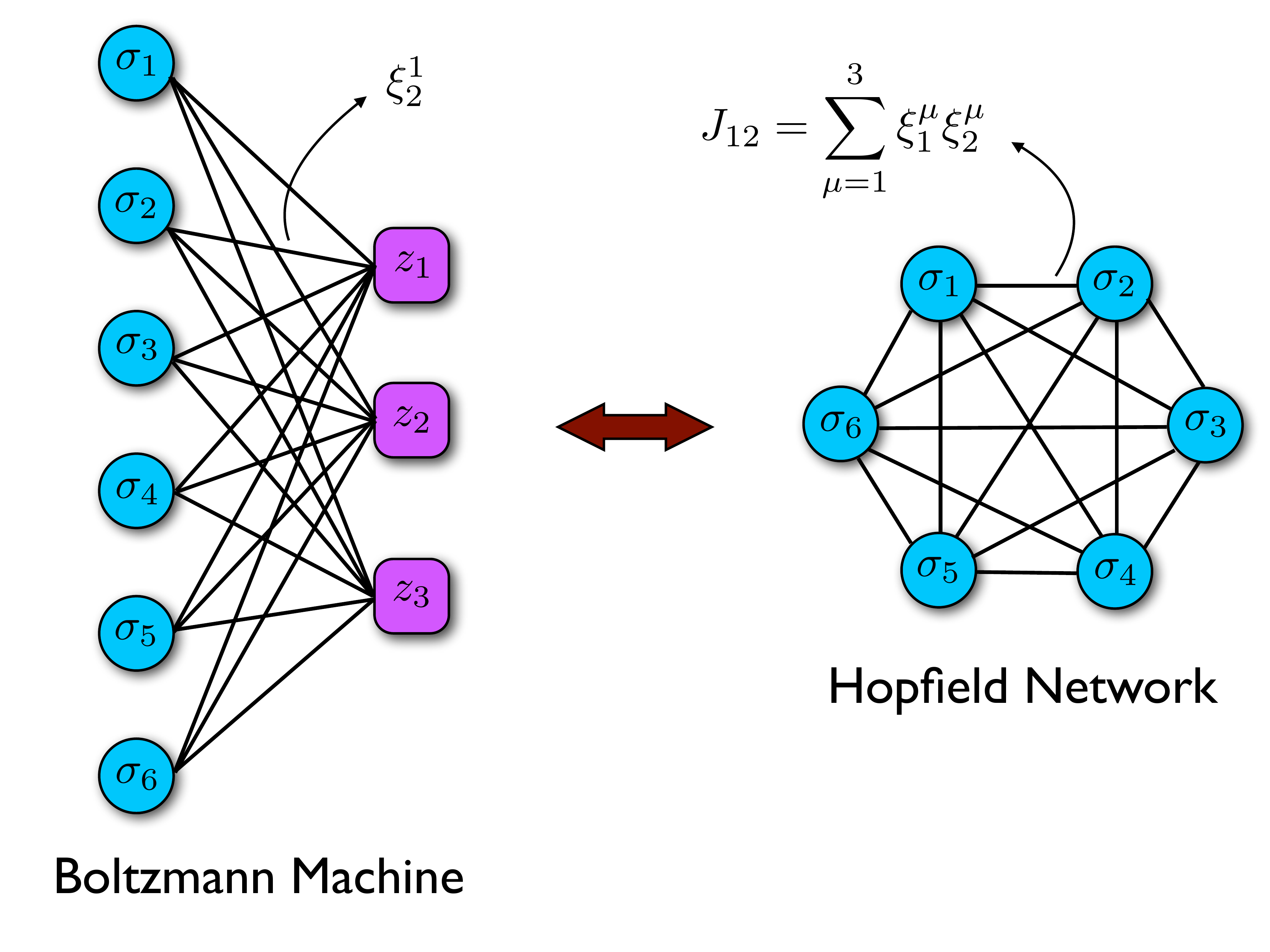}
\caption{Left panel: example of a RBM equipped with $6$ visible neurons $\sigma_i,\ i \in (1,...,6)$ and $3$ hidden units $z_{\mu},\ \mu \in (1,...,3)$. The weights connecting them form the $N \times P$ matrix $\xi_i^{\mu}$. Right panel: example of the corresponding AHN, whose six visible neurons $\sigma_i,\ i \in (1,...,6)$ retrieve as patterns stored in the Hebb matrix $J_{ij}=\sum_{\mu}^p \xi_i^{\mu}\xi_j^{\mu}$ the three vectors $\xi^{\mu},\ \mu \in (1,...,3)$, each pertaining to a {\em feature}, i.e. one of the three $z_{\mu}$ hidden variables of the (corresponding) RBM.} \label{fig:equivalenza}
\end{center}
\end{figure*}
Let us deepen the ideas exposed so far,  by introducing the standard definitions and concepts for Hopfield neural networks. Following classical  notations \cite{Ton}, we shall consider $N$ binary neurons (i.e., Ising spins \cite{AMIT}) and to each neuron $i$ we assign  a dichotomic variable $\si$ that describes its activity: if $\si = +1$ the $i$-th neuron is spiking, while if $\si = -1$ it is quiescent.
\newline
Neurons are embedded in a fully connected network, in such a way that \emph{mean-field} approaches are suitable for the investigation.
The synaptic potential $h_i$ that the $i$-th neuron receives from the other $N-1$ is defined as
\begin{equation*}
 h_i = \sum_{\substack{ j\neq i}}^N J_{ij} \sj \; ,
\end{equation*}
where $J_{ij}=J_{ji}$ is the synaptic coupling between neuron $j$ and neuron $i$, defined according to Hebb's learning rule \cite{Hebb} as
\begin{equation} \label{hebbrule}
 J_{ij} = \frac{1}{N} \summup \xiimu \xijmu \; .
\end{equation}
Indeed, associative memory models are built to recognize a certain group of words, pixels, or  generically {\em patterns} $\xi$: a \emph{pattern} is defined as a sequence of random variables $\xi = (\xi_1, \ldots, \xi_N)$.
If we want the network to memorize and retrieve a number $p$ of patterns, we have to introduce another index to distinguish them: $\{ \xi^1, \ldots, \xi^p\}$, and we shall assume that the set $\left\{ \xi_i^\mu \right\}_{i,\mu}$ is made of $p \times N$ i.i.d. variables. Notice that, for a Shannon information compression argument, if the network is able to cope with this kind of pattern, then it certainly retains at least the same  capacity in the case of correlated patterns \cite{correlated,ABDG}.
\newline
Boolean binary patterns have entries such that $\mathbb{P} (\xi_i=+1) = \mathbb{P}(\xi_i=-1) = 1/2$ , while Gaussian real patterns have entries drawn from $\mathbb{P}(\xi_i) \sim \vargauss$.
\begin{defi}
The Hamiltonian $H_N^{AHN} (\sigma,\xi)$ of the Associative Hopfield Network (AHN) equipped with $N$ Ising neurons $\sigma$ and $p$ patterns is defined as
\begin{equation} \label{HopfieldH}
 H_N^{AHN} (\sigma,\xi) = - \frac{1}{2N} \sum_{i,j}^N \summup  \xiimu \xijmu \si\sj \; .
\end{equation}
\end{defi}
Once introduced the (fast) noise $\beta=1/T \in \mathbb{R}^{+}$, where $T$ plays as a {\em temperature} in standard Statistical Mechanics, the partition function  $Z_{N,p}^{AHN} (\beta)$ for the AHN is defined as
\begin{equation*}
Z_{N,p}^{AHN} (\beta) = \sum_{\sigma} \exp \left\{ \frac{\beta}{2N} \summup \sum_{i,j}^N \xiimu \xijmu \si\sj \right\}.
\end{equation*}
and the free energy as $1/N \E_\xi \log Z_{N,p}^{AHN} (\beta)$, whose analysis allows inferring the model phase-diagram in the thermodynamic limit $(N \to \infty)$ \cite{AMIT}. Note that in the previous definitions we have introduced for simplicity also self-interactions, but we will see their presence doesn't affect the thermodynamic state of the network because they contribute at most to a simple constant term in the free energy.
\begin{defi}
The Hamiltonian $H_N^{RBM} (\sigma,\xi)$ of the Restricted Boltzmann Machine (RBM), equipped with a visible layer of $N$ binary (i.e. Boolean) units $\sigma_i$, $i \in (1,...,N)$ and a hidden layer of $p$ real (i.e. Gaussian) units $z_{\mu}$, $\mu \in (1,...,p)$, connected by the $N \times p$ weight matrix $\xi_i^{\mu}$, is defined as
\begin{equation}\label{BoltzmahhH}
 H_N^{RBM} (\sigma,\xi) = - \frac{1}{\sqrt{N}} \sum_{i,\mu}^{N,p} \xi_i^{\mu}\sigma_i z_{\mu}.
\end{equation}
\end{defi}
Again considering $\beta$  the fast noise of the network, the partition function $Z_{N,p}^{RBM} (\beta)$ for the RBM is introduced as
\begin{equation*}
Z_{N,p}^{RBM} (\beta) = \sum_{\sigma} \int_{\R^p} d\MG(z) \exp \left\{ \sqrt{\frac{\beta}{N}} \summup \sumi \xiimu \si \zmu \right\},
\end{equation*}
where $d\MG(z) = \prod_{\mu=1}^{p} \frac{dz_\mu}{\sqrt{2\pi}} e^{{z_\mu^2}/2}$ is the $p$-dimensional centered Gaussian measure. The model free-energy is defined as before.
It is just an exercise now to show (e.g., via standard Gaussian integration) the  following
\begin{proposition}
The partition functions of the Associative Hopfield Network and of the Restricted Boltzmann Machines are the same, i.e.
$$
Z_{N,p}^{AHN} (\beta) \equiv  Z_{N,p}^{RBM} (\beta),
$$
and thus the same equivalence holds for the two free energies.
\end{proposition}

Note that, while the identity $Z_N^{AHN} (\beta) \equiv  Z_N^{RBM} (\beta)$ holds only if we choose Gaussian hidden units $z_{\mu}$, an analogous equivalence can be proved introducing a class of generalised AHN and RBM models with any unit priors \cite{lettera,lungo}.

In order to investigate the capabilities of these networks to retrieve patterns, it is useful to introduce the concept of Mattis magnetization as follows.
\begin{defi}
For any $\mu \in (1,...,p)$, we define the Mattis magnetization, i.e. the overlap between the $\mu$-th patterns and the neuron states,  as
\begin{equation} \label{mattisoverlap}
 m_{\mu, N} (\sigma) = \frac{1}{N} \sumi \xiimu\si .
\end{equation}
In the following we will often drop the $N$ or $\s$ dependencies for lightening the notation.
\end{defi}
The magnitude of the Mattis magnetization encodes whether a pattern $\mu$ has been retrieved or not. Moreover we can rewrite the Hamiltonian \eqref{HopfieldH} as a function of the order parameters $m_\mu$'s as
\begin{equation*}
 H_N^{AHN}(\sigma,\xi) = - \frac{N}{2} \summup m_\mu^2 \; ,
\end{equation*}
hence it becomes clear, that its energy minima are located at large $m_\mu$. This means that the energy function is minimized as the spins are aligned to some of the $p$ patterns, thus indicating a retrieving state (i.e. the network overall works as a distributed associative memory).
\newline
Let us now turn our attention to the RBM case. Its energy function \eqref{BoltzmahhH} can be rewritten as
\begin{equation*}
 H_N^{RBM}(\sigma,\xi) = -  \sqrt{N} \sum_{\mu=1}^p m_{\mu}z_{\mu},
\end{equation*}
thus, if the system is in the retrieval region, i.e.,  there is some pattern $\mu$ (say $\mu^*$) that is retrieved by the Hopfield network, its related Mattis magnetization raises from zero acting as a {\em staggered magnetic field} over its related hidden variable $z_{\mu^*}$. In the Machine Learning perspective, this condition corresponds to selecting a {\em feature}, among the $p$ possible, and allows a statistically significant classification of the data.

\section{Mixed Hebbian networks} \label{hybrid}
In our ``hybrid'' Hopfield model, we consider the case in which the network has stored a low load of Boolean patterns and a high load of Gaussian ones. We will assign the variables $\tilde{\xi}^\nu$, $\nu = 1,\ldots, k = \gamma \ln N$ to the binary memories
and $\xi^\mu$, $\mu = 1, \ldots, p = \alpha N$ to the real ones (with  $\gamma, \alpha \geq 0$). We have
\begin{equation*}
\begin{cases}
 \mathbb{P} \{ \xiinu = +1\} = \mathbb{P} \{\xiinu = -1\} = \frac{1}{2} \quad & \forall i=1,\ldots,N \text{ and }
                              \nu=1,\ldots,k,\\
\mathbb{P} (\xiimu) \sim \vargauss & \forall i=1,\ldots, N \text{ and } \mu=1,\ldots,p \; .
\end{cases}
\end{equation*}
Following the description of the standard Hopfield neural network given in Section \ref{hnn}, we give the following
\begin{defi}
The Hamiltonian $H_N^{MHN} (\sigma, \xi, \xit)$ of the  mixed Hebbian network (MHN), equipped with $N$ Ising neurons, a low load of $k$ binary patterns and a high load of $p$ real patterns, reads as
\be
H_N^{MHN} (\sigma, \xi, \xit) = - \frac{1}{N} \sum_{1\leq i<j \leq N} \left( \sumnu \xiinu \xijnu + \summup \xiimu\xijmu \right) \si\sj.
\ee
\end{defi}
Notice that, splitting the above summations over $(i,j)$, the Hamiltonian of the mixed Hebbian network can be written as
\be\label{HNbis}
  H_N (\sigma, \xi, \xit)    =   - \frac{1}{2N} \sumij (\sumnu \xiinu \xijnu + \summup \xiimu\xijmu)  \si\sj + \frac{1}{2N} \sumi \summup \left( \xiimu \right)^2+ \frac{k}{2}  , \;
\ee
hence the last term at the r.h.s. of the previous equation does not contribute at all in the thermodynamic limit, while the second-last term converges to
$$
\lim_{N \to \infty}\left[\frac{1}{2N} \sumi \summup \left( \xiimu \right)^2\right] = \frac{\alpha}{2}.
$$
\begin{defi}
The Gibbs measure for a generic function of the neurons $F(\sigma)$ at a given level of noise $\beta$ is
\begin{equation} \label{Bmeasure}
  \omega_N(F)  = \frac{\sum_\sigma F(\sigma) e^{-\beta H_N(\sigma,\xi,\xit)}}{\ZN}.
\end{equation}
\end{defi}
Note that for $\beta \to 0$  the measure becomes flat, while for $\beta \to \infty$, i.e. at zero temperature,  only the global minima of the energy contribute to the measure.  
\begin{defi}
Given  $s$ independent  realizations (i.e., {\em replicas}) of the system, at the same noise level $1/ \beta$ and quenched patterns $\xi$ and $\xit$, we define the $s$-replicated Gibbs measure as $\Omega = \omega^1 \times \omega^2 \times \ldots \times \omega^s$, i.e. for any function of the $s$ neuron replicas $F(\sigma^{(1)},\ldots,\sigma^{(s)})$, 
\begin{align} \label{Bmesreplica}
\begin{split}
 \Omega \bigl(F(\sigma^{(1)},\ldots,\sigma^{(s)}) \bigr) =  \frac{1}{Z_N^s} \sum_{\sigma^{(1)}} \cdots \sum_{\sigma^{(s)}}
           F(\sigma^{(1)},\ldots,\sigma^{(s)})
            \exp\left\{ -\beta \sum_{a=1}^s H_N(\sigma^{(a)},\xi,\xit) \right\} \; .
  \end{split}
\end{align}
\end{defi}
\begin{defi}
The average over the quenched memories $\{ \xiinu \}_{i,\nu}$ and $\{ \xiimu \}_{i,\mu}$ for a generic function
$F(\xi,\xit)$ is introduced as
\begin{equation*}
 \E \lpq F(\xi,\tilde{\xi}) \rpq = \int \prod_{\mu = 1}^p \prod_{i=1}^N \frac{d\xiimu}{\sqrt{2\pi}} e^{- \frac{(\xiimu)^2}{2}}
                              \times \prod_{\nu=1}^k \prod_{j=1}^N \sum_{\{\xijnu\}} \frac{1}{2} F(\xi,\tilde{\xi}) \; .
\end{equation*}
Moreover we define the  average  $\lb \cdotp \rb= \E\Omega (\cdotp)$.
\end{defi}
We continue by introducing the  order parameters necessary to carry out the analysis of the mixed model. For any pattern, we define the Mattis magnetization as before for describing thermodynamic states in the retrieval phase, while we introduce  overlaps among replicas, as in \cite{BGG-JSP2010,BGGT-JSM2012} , for describing ordered states that are not correlated with patterns.
\begin{defi}
Given two  configurations $(a,b)$ of the network, the overlap $q_{ab}$ between visible units is defined as
\begin{equation} \label{overlapneuroni}
 q_{ab} (\boldsymbol{\s})= \frac{1}{N} \sum_{i=1}^N \sigma_i^{(a)} \sigma_i^{(b)}  \in [-1,1],
\end{equation}
and the overlap $p_{ab}$ between  hidden units as
\begin{equation}\label{overlapz}
 p_{ab}(\boldsymbol{z}) = \frac{1}{p} \sum_{\mu=1}^{p}  z_\mu^{(a)} z_\mu^{(b)} \in (-\infty,+\infty) .
\end{equation}
\end{defi}

Finally, we introduce the free energy density as 
\begin{defi}
We define the free-energy density $A(\alpha,\beta)$ of the mixed Hebbian network as
\be
A(\alpha,\beta)= \lim_{N \to \infty} A_{N,k,p}(\beta),\ \ \  A_{N,k,p}(\beta)=\frac1N \mathbb{E} \ln Z_{N,k,p}(\beta),
\ee
where the partition function $Z_{N,k,p}(\beta)$ reads as
\begin{eqnarray}  \label{ZNN}
Z_{N,k,p}(\beta)&=&\sum_{\sigma} \exp\{-\b H_N^{MHN} (\sigma, \xi, \xit)\}\nonumber\\
&=&\sum_{\sigma} \exp \left\{ \frac{\beta}{2N} \sumij \sumnu \xiinu\xijnu\si\sj + \frac{\beta}{2N} \sumij  \summup \xiimu\xijmu\si\sj \right\}.
\end{eqnarray}

%
\end{defi}

Therefore, the free-energy density at finite volume reads as
\begin{align} \label{preliminarypressure}
\begin{split}
  A_{N,k,p} & (\beta)  =  \frac{1}{N} \E \log Z_{N,k,p}(\beta) = \\
         = & \frac{1}{N} \E \Biggl[ - \frac{\beta k}{2} - \frac{\beta}{2N} \sumi \summup (\xiimu)^2 \Biggr] + \\
           & + \frac{1}{N} \E \log \Biggl( \sum_{\sigma} \exp \Biggl\{ \frac{\beta}{2N} \sumij \sumnu \xiinu\xijnu\si\sj
             + \frac{\beta}{2N} \sumij \summu \xiimu\xijmu\si\sj \Biggr\} \Biggl) \\
         = & - O \lp \frac{\ln N}{N} \rp - \frac{\alphaN \beta}{2} +  \\
           & + \frac{1}{N} \E \ln \Biggl( \sum_{\sigma} \exp \Biggl\{ \frac{\beta}{2N} \sumij \sumnu \xiinu\xijnu\si\sj
             + \frac{\beta}{2N} \sumij \summup \xiimu\xijmu\si\sj \Biggr\} \Biggl) \; ,
 \end{split}
\end{align}
in which the parameter $\alphaN$ is such that $\alphaN = \frac{p}{N} \to \alpha$ for $N\to \infty$.\\

We recall that, in the statistical mechanical treatment, finding an explicit expression for the free-energy density $A(\alpha,\beta)$ is the first step for understanding the properties of the network's thermodynamic states. This is because the solution of $A(\alpha,\beta)$ usually comes with a variational large deviation principle over the order parameters $\{ m_\mu,\ q_{ab},\ p_{ab}\}$.

\section{Sum rules for the mixed Hebbian network's free energy}\label{sum-rule}

In this Section we expose the interpolating structure that we set up to obtain an expression of the MHN free energy density, at the replica symmetric level, as a variational principle over the order parameters. The solution of this optimization problem is encoded  into a set of self-consistent equations that the order parameters have to satisfy, giving the phase diagram of the model by varying the external parameter.
\newline
In particular, the question we are addressing in the present work is about the existence of a retrieval phase in such a phase diagram: we will prove that there is actually a region in the $(\alpha,\beta)$ plane where the mixed Hebbian network is able to retrieve, in particular where the signal conveyed by the binary patterns is detectable over the real noisy sea.

Summarizing the strategy, we will first generalize the partition function (\ref{ZNN}) by letting it depend on three interpolating parameters, namely
$t \in \R^+$, $x \in \R^k$, $\psi \in [0,1]$ that, once set to proper values (i.e., $t=\beta$, $x=0$ and $\psi=1$), recovers the original
one of the mixed Hebbian network (see eq. \ref{ZNgen}). This interpolation will allow us to split the problem into two (related) sub-problems: one involving the Gaussian patterns, tackled by the stochastic stability technique in $\psi$, and the other involving the Boolean patterns, treated via the Hamilton-Jacobi technique in the $1+k$ dimensional space $(t,\ x)$.
\newline
Once formulated a sum rule for the free energy (see eq. \ref{tfc}), to set up the stochastic stability approach, we will introduce three external fields $\mathcal{A},\ \mathcal{B},\ \mathcal{C}$, where $\mathcal{A}$ acts on the $\{\sigma\}$ party, while $\mathcal{B}$ and $\mathcal{C}$ act on the $\{z\}$ party: while explicit expressions for these fields will be set a fortiori, their meaning can be discussed immediately. They are required to ensure that the interpolative procedure in $\psi$,  always reproduces the correct statistics on the neurons but in a mean field picture where units are no longer coupled.  The Hamilton-Jacobi formalism is naturally introduced when dealing with the explicit calculation of $A_{N,k,p} (t,x,\psi=0)$, which represents the free energy density of a Hopfield network with binary patterns and an external random field supplied by the Gaussian sea (that comes into play in terms of a quenched noise, hence against retrieval). In fact, $A_{N,k,p} (t,x,\psi=0)$ can be interpreted as the Guerra Action for a unitary-mass point-particle evolving in the $1+k$ dimensional $(t,\ x)$ space and can consequently be approached via standard techniques of Analytical Mechanics \cite{guerra-HJ,HJ-Jstat}.
\newline
We stress that the order in which we apply these two methods is interchangeable and in Appendix \ref{inverseprocess} we show
how, reasonably proceeding the other way around (that is, using first the Hamilton-Jacobi streaming and, later, the stochastic stability), we obtain the same results.

As a preliminary step, it is useful to apply the Gaussian integration to the partition function \eqref{ZNN} to linearize the Gaussian section of the free energy density function $A_{N,k,p}(\beta)$ with respect to the bilinear quenched memories carried by $\xiimu \xijmu$.
Namely:
\begin{align*}
\begin{split}
Z_{N,k,p}(\beta) = & \exp \Biggl\{ -\frac{\beta k}{2} + \frac{\beta}{2N} \sumi \summup (\xiimu)^2 \Biggr\} \times \\
            & \times \sum_{\sigma} \exp \left\{ \frac{\beta}{2N} \sumij \sumnu \xiinu\xijnu\si\sj + \frac{\beta}{2N} \sumij
               \summup \xiimu\xijmu\si\sj \right\} = \\
          = & \exp \Biggl\{ -\frac{\beta k}{2} + \frac{\beta}{2N} \sumi \summup (\xiimu)^2 \Biggr\}
             \sum_\sigma  \exp \left\{ \frac{\beta}{2N} \sumij \sumnu \xiinu\xijnu\si\sj \right\} \times \\
             & \times \int_{\R^p} d\MG(z) \exp \left\{ \sqrt{\frac{\beta}{N}} \summup \sumi \xiimu \si \zmu \right\} \; ,
\end{split}
\end{align*}
where $d\MG(z) = \prod_{\mu=1}^{p} \frac{dz_\mu}{\sqrt{2\pi}} e^{{z_\mu^2}/2}$ is the $p$-dimensional Gaussian measure.
\newline
As anticipated earlier, to achieve our goal we shall now analyse a generalized problem, for which we give hereafter the definition:
\begin{defi}
Once introduced $k+2$ scalar parameters $t \in \mathbb{R}^+,\ x \in \mathbb{R}^k, \psi \in [0,1]$, and three scalar fields $\mathcal{A},\ \mathcal{B},\ \mathcal{C}$, the generalized partition function $Z_N(t,x,\psi)$ for the mixed Hebbian network is defined as
\begin{align} \label{ZNgen}
\begin{split}
  Z_N & (t,x,\psi) = \exp \Biggl\{- \frac{\beta k }{2} - \frac{\beta}{2N} \sumi \summup (\xiimu)^2 \Biggr\} \times  \\
         & \times \sum_{\sigma} \int_{\mathbb{R}^p} d\MG(z) \
           \exp \Biggl\{ \frac{t}{2N} \sumij \sumnu \xiinu\xijnu\si\sj + \sumnu x_\nu \sumi \xiinu\si\Biggr\} \times \\
         & \times \exp \Biggl\{ \sqrt{\psi} \sqrt{\frac{\beta}{N}} \summup \sumi \xiimu\si\zmu \Biggr\} \times
           \exp \Biggl\{ A \sqrt{1-\psi} \sumi \ei\si \Biggr\}  \times \\
         & \times \exp \Biggl\{ B \sqrt{1-\psi} \summup \tmu \zmu \Biggr\}
           \times \exp \Biggl\{ C \frac{1-\psi}{2} \summup (\zmu)^2 \Biggr\} \; ,
\end{split}
\end{align}
with $\tmu,\ei \sim \vargauss$ $\forall \mu = 1,\ldots,p$, $i=1,\ldots,N$.
\end{defi}
Note that, by now, the scalar fields are given in full generality and they will be chosen later on, in order to ensure that the replica symmetric framework is preserved at the end of the interpolation.
\newline
Note further that, in perfect analogy we can extend also the free energy density function to $A_{N,k,p} (t,x, \psi)$, the Gibbs measures to $\omega_{t,x,\psi}$ and $\Omega_{t,x,\psi}$ and the overall average to $\lb \cdotp \rb_{t,x,\psi}$. Of course, also these quantities recover the standard statistical mechanical scenario once evaluated at $t=\beta$, $x=0$ and $\psi=1$.\\
\newline
\newline
We begin the study of the free energy density function through the stochastic stability. First, exploiting the Fundamental Theorem of Calculus
on $A_{N,k,p} (t,x, \psi)$ in the $\psi$ variable we write the next
\begin{proposition}
The following sum rule for the generalised free energy  $A_{N,k,p} (t,x, \psi)$ of the mixed Hebbian network holds
\begin{equation} \label{tfc}
 A_{N,k,p} (t,x) = A_{N,k,p} (t,x, \psi = 1) = A_{N,k,p} (t,x,\psi = 0) + \int_0^1 \left(\partial_{\psi'} A_{N,k,p}(t,x,\psi')\right)_{\psi'=\psi} d\psi \; .
\end{equation}
\end{proposition}
The original problem is therefore recast in the evaluation of the two terms at the r.h.s. of eq. (\ref{tfc}).
\newline
To compute the first term we start through a standard Gaussian integration, hence
\begin{align}  \label{alpha0}
\begin{split}
 A_{N,k,p} & (t,x, \psi=0) = \diagonale + \\
          & + \frac{1}{N} \E \Biggl[ \log  \sum_{\sigma} \exp \Biggl\{ \frac{t}{2N}
            \sumij \sumnu \xiinu\xijnu\si\sj + \sumnu x_\nu \sumi \xiinu\si + \mathcal{A} \sumi \ei \si \Biggr\} \times \\
          & \times \int_{\mathbb{R}^p} \frac{dz_1 \cdots dz_p}{(2\pi)^{p/2}} \exp \Biggl\{ \summup \biggl( \mathcal{B} \tmu \zmu +
            \frac{\mathcal{C}-1}{2} \zmu^2 \biggr) \Biggr\} \Biggr] = \\
        = & \diagonale + \frac{1}{N} \E \ln \Biggl( \frac{1}{(1-\mathcal{C})^{p/2}} e^{\frac{\mathcal{B}^2 \theta^2}{2(1-\mathcal{C})} p} \Biggr) + \\
          & + \frac{1}{N} \E \ln \sum_\sigma \exp \Biggl\{ \frac{t}{2N} \sumij \sumnu \xiinu\xijnu \si\sj +
            \sumnu x_\nu \sumi \xiinu \si + \mathcal{A}\sumi \ei\si \Biggr\} \; .
\end{split}
\end{align}
It is now crucial to notice that the fourth term of Eq.~\eqref{alpha0} can be interpreted as the free energy density $\tilde{A}_{N,k}(t,x)$ of a Hopfield network with $k$ binary patterns $\{ \tilde{\xi}^\nu \}$ and $N$ external random fields $\mathcal{A} \eta_i$: note that the latter account for the slow noise supplied by the underlying sea of Gaussian patterns that can not be retrieved.

It is convenient to rename this free energy density $\tilde{A}_{N,k}(t,x)$ by the following definition:
\begin{defi}
Once introduced a generalized partition function $Z_{N,k}(t,x)$, identified by the following expression
\begin{equation*}
Z_{N,k}(t,x) = \sum_\sigma \exp \Biggl\{ \frac{tN}{2} \sumnu \mnu^2 + N \sumnu x_\nu \mnu + \mathcal{A} \sumi \ei\si \Biggr\} \; ,
\end{equation*}
we define the Guerra Action $\tilde{G}_{N,k}(t,x)$, for a unitary-mass point-particle moving in the $(1+k)$ dimensional $(t, x)$ space, as the negative free energy density $\tilde{A}_N(t,x)$:
\be
\tilde{G}_{N,k}(t,x) = -\tilde{A}_{N,k}(t,x) = - \frac1N \ln \Zt.
\ee
\end{defi}
With this definition, the application of the Hamilton-Jacobi formalism for handling $\tilde{A}_{N,k}(t,x)$ is straightforward. In fact, it is immediate to check that, as $\tilde{A}_{N,k}(t,x)$ has the following properties
\begin{align}  \label{propalpha}
 \partial_t \tilde{A}_{N,k}(t,x) &= \frac{1}{2} \sumnu \lb \mnu^2 \rb_{x,t} & \partial_{x_\nu} \tilde{A}_{N,k}(t,x) &= \lb \mnu \rb_{x,t} \; ,
\end{align}
we can proceed according to the Hamilton-Jacobi prescription for $\tilde{G}_{N,k}(t,x)$. In fact, thanks to the properties  ~\eqref{propalpha}, it is
immediate to verify the next
\begin{proposition}
The Guerra Action obeys the following Hamilton-Jacobi streaming
\begin{equation} \label{HJ}
 \partial_t \bigl(\tilde{G}_{N,k}(t,x) \bigr) + \frac{1}{2} \bigl( \partial_x \tilde{G}_{N,k}(t,x) \bigr)^2 + V_{N,k} (t,x) = 0 \; ,
\end{equation}
where the potential $V_{N,k}(t,x)$ is given by the sum over all the binary patterns of their related Mattis magnetization's variances, namely
$$
V_{N,k}(t,x) = \frac{1}{2} \sum_\nu^k \bigl( \lb \mnu^2 \rb_{t,x}  - \lb \mnu \rb^2_{t,x} \bigr) =
\frac{1}{2N} \partial^2_{xx} \tilde{G}_{N,k}(t,x).
$$
\end{proposition}
\begin{remark}
As we are in the low-storage regime for binary patterns (i.e., $k \propto \ln N$), in the thermodynamic limit  the Guerra  Action paints a Galilean trajectory for the point-like particle: its evolution is simply a free motion as $\lim_{N\to\infty} V_{N,k}(t,x)=0$.
\end{remark}
\begin{proposition}
If we define a $k$-dimensional vector $\Gamma_N (t,x)$, whose components are $\Gamma_N^\nu (t,x) =  \partial_{\xnu} \tilde{G}_{N,k}(t,x)$, by deriving Eq.~\eqref{HJ} with respect to $\xnu$ we obtain the following set of $k$ Burgers equations for the canonical momenta
\begin{equation} \label{gammaN}
 \partial_t \Gamma_N^\nu (t,x) + \sum_{\tau=1}^k \Gamma_N^\tau (t,x) \times \partial_{x_\tau} \Gamma_N^\nu (t,x) =
   \frac{1}{2N} \sum_{\tau=1}^k \partial^2_{x_\tau x_\tau} \Gamma_N^\nu (t,x) \quad \forall \nu \; .
\end{equation}
\end{proposition}
At present, the goal is thus to solve the Burgers equations and integrate back the solutions to get the original problem for $\tilde{G}_{N,k}(t,x)$ (and therefore for $\tilde{A}_{N}(t,x)$) solved too. As standard, performing the Cole-Hopf transform $\Phi_{N,k} (t,x) := e^{N \tilde{A}_{N,k}(t,x)}$, we can assert that
\begin{proposition}
Solving expression ~\eqref{gammaN} is equal to solve the following Cauchy problem for the heat equation
\begin{equation}  \label{calore}
 \begin{cases}
  \partial \Phi_{N,k} (t,x) - \frac{1}{2N} \varDelta \Phi_{N,k} (t,x) = 0  & t \in \R, x \in \R^k, \\
  \Phi_{N,k} (0,x) = e^{N \tilde{A}_{N,k} (0,x)}  & x\in\R^k \; .
 \end{cases}
\end{equation}
\end{proposition}
We can now deal with the problem above through standard techniques. Namely we write
\begin{equation}   \label{solcalore}
 \Phi_{N,k} (t,x) = \int_{\R^k} dx'_1 \cdots dx'_k G(t,x-x') \Phi_{N,k} (0,x') \; ,
\end{equation}
where $G$ is the Green propagator $G(t,x) = \lp \frac{N}{2\pi t} \rp^{k/2} e^{- \frac{\sum_\nu \xnu^2 N}{2t}}$. \\
The computations for the initial condition $\Phi_{N,k} (0,x)$ return
\be
  \Phi_N (0,x)  = \exp \Biggl\{ N\ln2 + \sumi \E \ln\cosh \biggl( \sumnu \xiinu\xnu + A\eta \biggr) \Biggr\} \; .
\ee
Therefore, we can state that
\begin{theorem}
The solution to the problem in~\eqref{calore} is given by the following saddle point equation:
\begin{equation*}
 \Phi_{N,k} (t,x) = \lp \frac{N}{2\pi t} \rp^{k/2} \int_{\R^k} dx'_1 \cdots dx'_k \ e^{-N g(t,x,x')} \; ,
\end{equation*}
\begin{equation} \label{g}
 g(t,x,x') = \frac{1}{2t} \sumnu (\xnu - \xnu')^2 - \ln2 - \frac{1}{N} \sumi \E \ln \cosh \biggl( \sumnu \xiinu \xnu'
             + A\ei \biggr) \; .
\end{equation}
\end{theorem}
\begin{corollario}
Recalling that $\tilde{A}_{N,k}(t,x)  = \frac{1}{N} \ln \Phi_N (t,x)$, in the thermodynamic limit we have that
\begin{equation} \label{HJresult}
 \tilde{A} (t,x) = \lim_{N \to +\infty} \tilde{A}_{N,k}(t,x) = - \min_{x'\in\R^k} g(t,x,x') \; .
\end{equation}
\end{corollario}
To get the full expression of the Guerra Action  in the thermodynamic limit, we must finally set
$t = \beta$, $x = 0$ and perform  the minimization of the function $g$ given in~\eqref{g}: with these values for $t$ and $x$, we have to fix $\xnu' = \beta \lb \mnu \rb $ $\forall \nu=1,\ldots,k$.

At this point equation~\eqref{tfc} is almost all explicit.
We still need to calculate the integral term at the top right side of equation~\eqref{tfc}, for which it is
sufficient to evaluate the $\psi$-derivative of the free-energy density $A_{N,k,p} (t,x, \psi)$ and write it in a way that allows to extrapolate easily its replica symmetric approximation.
\newline
Here we just provide the final result, while the step-by-step calculations for the $\psi$-derivative
are left for the reader in Appendix \ref{noia}. So briefly,
\begin{align} \label{ss}
\begin{split}
 \frac{d A_{N,k,p} (t,x, \psi)}{d \psi} = & \frac{1}{N} \E \left[ \frac{d_\psi Z_{N,k,p} (t,x, \psi)}{Z_{N,k,p} (t,x, \psi)} \right] = \frac{1}{2N} \lp \beta - \mathcal{B}^2 - \mathcal{C} \rp \summup
                        \E\omega \lp \zmu^2 \rp_{t,x} + \\
                      & - \frac{\alphaN \beta}{2} \lb q_{12} p_{12} \rb_{t,x} - \frac{\mathcal{A}^2}{2} \lp 1 - \lb q_{12} \rb_{t,x}
                          \rp + \frac{\alphaN\beta^2}{2} \lb p_{12} \rb_{t,x} \; .
 \end{split}
\end{align}
Fixing the free parameters $\mathcal{A}$, $\mathcal{B}$ and $\mathcal{C}$ as
\begin{align} \label{fixparametri}
\mathcal{A} &= \sqrt{\alpha \beta \bar{p}} \; , & \mathcal{B} &= \sqrt{\beta \bar{q}} \; , & \mathcal{C} &= \beta(1 - \bar{q}) \; ,
\end{align}
and adding and subtracting the term $(\alphaN\beta \cdot \qb\pb)/2$ in Eq.~\eqref{ss} we have
\begin{equation} \label{ssfin}
 \frac{d A_{N,k,p} (t,x, \psi)}{d \psi}  = - \frac{\alphaN\beta}{2} \pb(1-\qb)  -\frac{\alphaN\beta}{2} \lb (q_{12} - \qb)(p_{12} - \pb) \rb_{t,x},
\end{equation}
In the replica symmetric regime, the order parameters $m$, $q_{12}$, $p_{12}$ do not fluctuate with respect
to their quenched averages in the thermodynamic limit, i.e. using a bar to denote their averages, $\lb m \rb_{t,x} \to \bar{m}$, $\lb q_{12}
\rb_{t,x} \to \bar{q}_{12} $, $\lb p_{12} \rb_{t,x} \to \bar{p}_{12} $ as $N \to \infty$.  By choosing $ \bar{p}=\bar{p}_{12} $ and $\bar{q}=\bar{q}_{12}$  the last term at the r.h.s. of the above expression goes to zero in the thermodynamic limit and the $\psi$-derivative can be integrated being constant over $\psi$.   It holds  \cite{bg,Bip,zigg,pizzo} that the optimal values of $\bar{p}$ and $\bar{q}$ can simply be obtained by computing the two overlaps at $\psi=0$ and this turns out to be equivalent to take the extremum of the trial free energy $(\ref{tfc})$ w.r.t. $\bar{p}$ and $\bar{q}$
as stated in the following main theorem.
\begin{theorem}\label{th:1}
 The replica-symmetric free-energy density of the mixed Hebbian network defined by the Hamiltonian ~\eqref{HNbis}, in the thermodynamic limit, is determined by extremizing $A(\boldsymbol{m},\bar{q},\bar{p};\alpha,\beta)$, where
 \begin{align} \label{Alpha}
 \begin{split}
  A(\boldsymbol{m},\bar{q},\bar{p};\alpha,\beta) = & - \frac{\alpha\beta}{2} - \frac{\alpha}{2} \ln \bigl( 1 - \beta (1 - \qb) \bigr)
          + \frac{\alpha\beta\qb}{2 \bigl(1 - \beta (1 - \qb)\bigr)} - \frac{\beta}{2} \sum_\nu \mnu^2 + \\
        & + \ln 2 + \left\lb \ln\cosh \biggl( \beta \sum_\nu \tilde{\xi}^\nu \mnu
          + \sqrt{\alpha\beta \pb} \eta \biggr) \right\rb - \frac{\alpha\beta}{2} \pb (1 - \qb) \; ,
 \end{split}
 \end{align}
 with $\eta\sim\vargauss$ and where the values of its order parameters are set via their following self-consistencies
\begin{eqnarray}
\pb  &=& \frac{\beta \qb}{\bigl( 1 - \beta (1 - \qb)\bigr)^2} \; , \label{selfp} \\
 \qb  &=& \left\lb \tanh^2 \biggl( \beta \sumnu \tilde{\xi}^\nu \mnu+ \sqrt{\alpha\beta \pb} \eta \biggr) \right\rb \; ,
         \label{selfq}  \\
 \mnu  &=&  \left\lb \tilde{\xi}^\nu \tanh \biggl( \beta \sumnu \tilde{\xi}^\nu \mnu + \sqrt{\alpha\beta\pb}\eta
                \biggr)  \right\rb \; . \label{selfm}
\end{eqnarray}
\end{theorem}
\begin{remark}
We highlight that for $\alpha = 0$ and $\nu = 1$ we recover the Curie-Weiss free energy density \cite{barra0}, while, if $\alpha > 0$ and $\nu = 0$ we recover the free energy density of the analog Hopfield model at high storage \cite{BGG-JSP2010} and, finally, keeping $\nu=0$, with $\alpha \to \infty$  (such that $\alpha \beta^2 = \beta'$, with $\beta'$ finite), we recover the expression of  the Sherrington-Kirkpatrick free energy density at noise level $\beta'$ \cite{BGGT-JSM2012,Bip}.
\end{remark}
\begin{remark}
In order to get insights in the critical behavior exhibited by the system, in the expression ~\eqref{selfq}, as standard when dealing with second-order phase transition, we can expand for small $q$ 
\begin{equation*}
 q \simeq \frac{\beta^2 \alpha}{(1-\beta)^2}q + o(q).
\end{equation*}
This procedure returns a (second order) transition line for ergodicity breaking at
\begin{equation*}
 \frac{\beta^2 \alpha}{(1-\beta)^2} = 1 \quad \Leftrightarrow \quad \beta = \frac{1}{1+\sqrt{\alpha}},
\end{equation*}
that is the same as the one for the (standard, i.e. digital) Hopfield network \cite{ags1,ags2} as well as for its analog counterpart \cite{BGG-JSP2010,BGGT-JSM2012}: this is not particularly surprising as we are checking here the pure ergodic/spin-glass transition where Universality is expected to hold \cite{carmona,genovese}.
\end{remark}
A different intuition is needed when searching the boundary (i.e. the transition line) splitting the spin-glass phase (whose existence has never been discussed) from a (possible) region of retrieval (whose existence is not straightforward).
\newline
To find this first-order transition line we must compare the values of the two free-energies (the one under the \emph{pure state} ansatz holding for retrieval and the other for no net magnetization accounting for the spin glass phase), check that there is a region in the $(\alpha, \beta)$ plane where one prevails over the other and a complementary region where the opposite is true. The transition line is just given by the set of points in the parameter space where the two free energy balance. Our results return the same transition (hence the same retrieval region) of the standard (i.e. digital) Hopfield network. Its analog counterpart does not retrieve at all hence there is no line to compare that case.
\newline
The whole can be restated in the following
\begin{proposition}
The mixed Hebbian network, equipped with an extensive load of real patterns and with a low load of binary patterns, is able to handle the binary patterns as long as the system stays confined within the standard retrieval region \cite{Ton}.
\end{proposition}

\begin{remark}
Once we fixed the parameters $\mathcal{A}$, $\mathcal{B}$ and $\mathcal{C}$ (and, in particular, noting that $\mathcal{A} = \sqrt{\alpha \beta \bar{p}}$) and we have an explicit expression for the mixed Hebbian network's free energy density (see eq. \ref{Alpha}), via its $\langle m_{\nu}\rangle$ self-consistency we can appreciate how the high load of real patterns acts as a disturbing noise against the signal carried by the booleans
\end{remark}
\begin{remark}
Note that, in the $\alpha \to 0$ limit (hence neglecting the real sea), the critical point becomes $\beta_c = 1$. This is perfectly consistent with the emergence of a ferromagnetic phase (i.e., the point $(\beta=1, \alpha=0)$ is the Curie-Weiss or Mattis critical point).
\end{remark}

Note that a fairly standard usage of the replica-trick allows to extend the previous result to the case of a high load of boolean patterns too. Since it is not  a rigorous argument we state the following as a   
\begin{conj}
Assuming an high storage of both real patterns (hence $p = \alpha N$) as well as binary patterns (hence $k = \gamma N$),  Theorem \ref{th:1}  keeps holding as long as we replace $\alpha \to \alpha + \gamma$.
\end{conj}

\section{Conclusions}
The Hopfield neural network and the restricted Boltzmann machine are amongst the best known and intensively studied models in Artificial Intelligence. The former is meant to mimic retrieval, namely the capacity of (the {\em neurons} of) a machine to recall a pattern of information previously stored. The latter is meant to mimic learning, namely the capacity of (the {\em synapses} of) a machine to be trained to encode selected patterns of information.
Remarkably, Hopfield networks and Boltzmann machines share the same thermodynamics. This equivalence has several implications and, in particular, it implies that  the conditions under which the former is able to retrieve are the same conditions under which the latter is able to identify features in the input.
In fact, in this equivalence, the patterns of information retrieved by the Hopfield model corresponds to the optimized weights of the trained Boltzmann machine.
\newline
However, in the wide Literature concerning these models, the patterns handled by the Hopfield model are typically binary, while the weights the Boltzmann Machine usually ends up with are real: this gap looks structural since the retrieval of real patterns (at least in the high-load regime) is beyond the Hopfield model capabilities. 
While numerical understanding in the field increases at an impressive rate, analytical improvements proceed more slowly. In order to get further insights into this point through the analytic perspective, in this work we considered a mixed Hopfield network, where patterns are partly real and partly binary and we studied its statistical mechanical properties (i.e., we focused on the behavior of {\em averaged systems} and {\em in the thermodynamic limit}, which is not the typical benchmark in Computer Science).
\newline
In particular, we rigorously answered (positively) to the question of whether such a hybrid network with a high-load of analog patterns and a low-load of binary patterns is able to retrieve the latter (on the other hand, the retrieval of a high-load of analog patterns is already known to be unfeasible \cite{bov-stoc,lungo}). We proved that the hybrid model shares the same phase diagram of the classic Hopfield network with a high storage of Boolean patters only: in the parameter space, where parameters are given by the fast noise (i.e., the temperature) and by the slow-noise (i.e., the ``sea'' of analog patterns), there exists a retrieval region bounded by a first-order transition line.

This result has been achieved by developing a novel interpolating technique entirely stemming from the Guerra scheme (see \cite{BGG-JSP2010,BGGT-JSM2012} and \cite{HJ-Jstat,Bip}). In a nutshell, exploiting the above mentioned equivalence, we recast the hybrid Hopfield model in terms of its related Boltzmann machine and then we ask for stochastic stability of the bulk of patterns (hence the real ones). We  interpolate between the free energy of the mixed Hopfield model and two one-body random systems (whose factorized treatment becomes straightforward). This approach allows us to recognize, within the free energy contribution due to real patterns, another nestling free-energy density due to the Boolean contribution of the binary patterns.  The latter can then be extracted via the Hamilton-Jacobi route in terms of its natural order parameters. This approach allows detecting when the signal carried by a logarithmic load of Booleans is strong enough to shine over the noisy sea generated by the extensive storage of Gaussian patterns.
\newline
Finally, we stress that this machinery does not apply in the case of an high load of real as well as binary patterns. This challenging case can however be addressed via a fairly standard replica-trick calculation obtaining evidence that the outlined scenario is preserved as long as the sum of the two slow noises (stemming from the two contributions of real and binary patterns) does not exceed the usual threshold.

\vspace{1cm}
{\bf Acknowledgements}

\noindent
E.A.  acknowledges financial support from GNFM-INdAM (Progetto Giovani Agliari-2016).\\
A.B. acknowledges financial support from Salento University and by GNFM-INdAM.\\
D.T. acknowledges financial support from GNFM-INdAM (Progetto Giovani Tantari-2016).\\


\appendix

\section{The inverse process}  \label{inverseprocess}

In this appendix we shall illustrate that proceeding first with the HJ formalism and then with the stochastic stability is
equivalent to the process we described in Sec.~\ref{sum-rule}. Briefly, the method consists of the following steps.
\newline
Now, instead of the generalized partition function defined in~\eqref{ZNgen}, we have the following:
\begin{align*} 
 Z_{N,k,p}(t,x) = & \sum_\sigma \exp \Biggl\{ \frac{t}{2N} \sumij \sumnu \xiinu\xijnu \si\sj + \sumnu \xnu \sumi \xiinu\si \Biggr\}
              \times \\
            & \times \exp \Biggl\{ -\frac{k\beta}{2} - \frac{\beta}{2N} \sumi \summup (\xiimu)^2 +
              \frac{\beta}{2N} \sumij \summup \xiimu\xijmu \si\sj  \Biggr\} \; ,
\end{align*}
where we can notice the Hamilton-Jacobi scaffold in the interpolation of the Boolean section of the system. We recover the proper partition function
if we put $t=\beta$ and $x=0$, while if $t=0$ and $x=1$ we obtain a one-body problem for the boolean memories. \\
Even though the generalized free energy is now defined through this new partition function, the equations for its derivatives expressed in~\eqref{propalpha}
still hold and therefore we can proceed with the Hamilton-Jacobi formalism adopting the same argument we used in Sec.~\ref{sum-rule}. So Eqs.~\eqref{HJ},~\eqref{gammaN} and~\eqref{calore} still hold, but now the initial state function $A_{N,k,p} (0,x)$ is
\begin{align*}
  A_{N,k,p}(0,x) = & \frac{1}{N} \E \ln \Biggl( \exp \Biggl\{ -\frac{k\beta}{2} - \frac{\beta}{2N} \sumi \summup (\xiimu)^2
                    \Biggr\} \sum_\sigma \exp \Biggl\{ \sumnu \xnu \sumi \xiinu\si \Biggr\}  \times \\
                  & \times \exp \Biggl\{ \frac{\beta}{2N} \sumij \summup \xiimu\xijmu\si\sj \Biggr\}  \Biggr) \; .
\end{align*}
This function is now interpretable as the free energy density at a finite volume $N$ of a Hopfield network with $p$ real
patterns and an external field (that this time contains patterns of information), so we can now use the stochastic stability technique to write an explicit form of the expression
above. To do so, we introduce the variable $\psi \in [0,1]$ and the interpolated free energy density:
\begin{align*}
 A_{N,k,p} (0,x,\psi) = & \diagonale + \frac{1}{N} \E \ln \Biggl( \sum_\sigma \exp \Biggl\{ \sum_\nu \sum_i \xnu\xiinu\si
                       \Biggr\} \times  \\
              & \times \int_{\R^p} Dz \exp \Biggl\{ \sqrt{\psi} \sqrt{\frac{\beta}{N}} \summup\sumi\xiimu\si\zmu \Biggr\} \times
                       \exp \Biggl\{ \mathcal{A} \sqrt{1 -\psi} \sumi \ei\si \Biggr\} \times \\
              & \times \exp \Biggl\{ \mathcal{B} \sqrt{1-\psi} \summup \theta_\mu \zmu \Biggr\}
                \times \exp \Biggl\{ \mathcal{C} \frac{1-\psi}{2} \summup \zmu^2 \Biggr\} \Biggr) \; .
\end{align*}
Mirroring the exposition reported in the main text,  we can now apply the Fundamental Theorem of Calculus in $\psi$, perform almost the same calculations and substitute the values of the free parameters according to~\eqref{fixparametri}. What we obtain is:
\begin{align*}
 A_{N,k,p} & (0,x) = A_{N,k,p} (0,x,\psi = 1) = A_{N,k,p} (0,x,\psi=0) + \int_0^1 d \psi \ (d_\psi' A_{N,k,p}(0,x,\psi'))_{\psi'=\psi} = \\
                = & \diagonale + \ln 2 + \frac{1}{N} \sumi \E \ln\cosh \Biggl( \sumnu \xiinu + \sqrt{\alphaN\beta\pb}\ \ei
                  \Biggr) + \\
                  & - \frac{\alphaN}{2} \ln \bigl( 1 - \beta(1-\qb) \bigr) +
                  \frac{\alphaN\beta\qb}{2 \bigl( 1-\beta(1-\qb) \bigr)} - \frac{\alphaN\beta}{2} \pb (1-\qb) \; .
\end{align*}
Now recalling that the solution to~\eqref{calore} is defined by~\eqref{solcalore}, and that $\Phi_{N,k,p} = e^{N A_{N,k,p}}$ we
can write the free energy density function at a finite volume $N$:
\begin{equation*}
 A_{N,k,p} (t,x) = \diagonale + \frac{1}{N} \ln \left( \frac{N}{2\pi t} \right)^{k/2} + \frac{1}{N} \ln \int_{\R^k}
                  e^{-Ng(t,x,x')} \; ,
\end{equation*}
where
\begin{align*}
g(t,x,x') = & \sumi \frac{ \left( x_i - x'_i \right)^2}{2t} - \ln2 - \frac{1}{N} \sumi \E \ln\cosh \left( \sumnu\xiinu\xnu +
              \sqrt{\alphaN\beta\pb}\ \ei \right) + \\
            & + \frac{\alphaN}{2} \ln \left( 1 - \beta(1-\qb) \right) - \frac{ \alphaN \beta\qb}{2 \bigl( 1-\beta(1-\qb)
              \bigr)} + \frac{\alphaN\beta}{2}\pb(1-\qb) \; .
\end{align*}
In the thermodynamic limit the free-energy density is consequently obtained by~\eqref{HJresult} with the help of a saddle point
argument. So, fixing the parameters $t$ and $x$ to be $t=\beta$, $x = 0$ and finding that the minimum of the
function $g$ is determined by $\xnu' = \beta \lb \mnu \rb$, we can write the following expression for $A(\alpha,\beta)$:
\begin{align*}
A(\alpha,\beta) = & -\frac{\alpha\beta}{2} + \ln 2 - \frac{\beta}{2} \sum_\nu \lb \mnu \rb^2
                + \Biggl\lb \ln\cosh \Biggl( \beta\sum_\nu \tilde{\xi}_\nu \lb\mnu\rb + \sqrt{\alpha\beta\pb}\eta \Biggr)
                  \Biggr\rb + \\
                & - \frac{\alpha}{2} \ln \bigl( 1- \beta(1-\qb) \bigr) + \frac{\alpha\beta\qb}{2\bigl( 1-\beta(1-\qb)
                  \bigr)} - \frac{\alpha\beta}{2}\pb(1-\qb) \; ,
\end{align*}
which is exactly the same as Eq.~\eqref{Alpha} that we found through the calculations of Sec.~\ref{sum-rule} where the order of
the methods were reverted.

\section{Calculating the $\psi$-streaming of the interpolating free energy}  \label{noia}

As anticipated in Sec.~\ref{sum-rule}, in this appendix we will illustrate the calculations of the $\psi$-derivative of the generalized free energy density $A_{N,k,p}(t,x,\psi)$ written in Eq.~\eqref{ss}.

When evaluating the streaming $d_\psi A_{N,k,p}(t,x,\psi)$ we get the sum of
four terms: $I$, $II$, $III$ and $IV$, that we shall analyse shortly. Each one comes as a consequence of the derivation of a corresponding exponential
term appearing into the expression of the generalized free energy density, whose generalized partition function $Z_{N,k,p}(t,x,\psi)$ is defined in~\eqref{ZNgen}. \\
We remind that we introduced in Sec.~\ref{sum-rule} the generalized average $\langle \cdot \rangle_{t,x,\psi}$, that naturally extends the Gibbs measure encoded in the
interpolating scheme (and is reduced to the proper one whenever setting $t=\beta$, $x=0$ and $\psi=1$). To lighten the expressions, we introduce the function $B_{N,k,p} (t,x,\psi)$ that stands for the generalized Boltzmann factor.  \\
We can now show the calculations of terms $I$, $II$, $III$ and $IV$:

\begin{align}
  I & = \frac{1}{N} \E \Biggl[ \sum_\sigma \int d\MG(z)\ \sqrt{\frac{\beta}{N}} \sumi \summup \xiimu \si \zmu \times
          \frac{1}{2\sqrt{\psi}} B_{N,k,p}(t,x,\psi) \Biggr] = \\
      & = \frac{\sqrt{\beta}}{2N\sqrt{N\psi}} \sumi \summup \E \Biggl[ \xiimu \omega_{t,x,\psi}(\si\zmu) \Biggr] = \\
      & = \frac{\sqrt{\beta}}{2N\sqrt{N\psi}} \sumi \summup \E \Biggl[ \partial_{\xiimu} \omega_{t,x,\psi}(\si\zmu) \Biggr] = \\
      & = \frac{\beta}{2N} \summup \E \omega_{t,x,\psi}(\zmu^2) - \frac{\alphaN\beta}{2} \lb q_{12} p_{12} \rb_{t,x,\psi} \label{I}
 \end{align}\\
\begin{align}
  II & = \frac{1}{N} \E \Biggl[ \frac{1}{\Ztx} \sum_\sigma \int d\MG(z)\ \frac{-\mathcal{A}}{2\sqrt{1-\psi}} \sumi \ei\si B_{N,k,p}(t,x,\psi) \Biggr] = \\
     & = -\frac{\mathcal{A}}{2N\sqrt{1-\psi}} \sumi \E \Biggl[ \ei \omega_{t,x,\psi} (\si) \Biggr] = \\
     & = -\frac{\mathcal{A}}{2N\sqrt{1-\psi}} \sumi \E \Biggl[ \partial_{\ei} \omega_{t,x,\psi}(\si) \Biggr] = \\
     & = -\frac{\mathcal{A}^2}{2} \bigl( 1 - \lb q_{12} \rb_{t,x,\psi} \bigr) \label{II}
 \end{align}\\
%
\begin{align}
  III & = \frac{1}{N} \E \Biggl[ \frac{1}{\Ztx} \sum_\sigma \int d\MG (z)\ \frac{-\mathcal{B}}{2\sqrt{1-\psi}} \summup \tmu \zmu B_{N,k,p}(t,x,\psi)
        \Biggr] = \\
    & = -\frac{\mathcal{B}}{2N\sqrt{1-\psi}} \summup \E \Biggl[ \tmu \omega_{t,x,\psi} (\zmu) \Biggr] = \\
    & = -\frac{\mathcal{B}}{2N\sqrt{1-\psi}}
        \E \Biggl[ \partial_{\tmu} \omega_{t,x,\psi}(\zmu) \Biggr] = \\
    & = -\frac{\mathcal{B}^2}{2N} \summup \E \omega_{t,x,\psi} (\zmu^2) + \frac{\alphaN \mathcal{B}^2}{2} \lb p_{12} \rb_{t,x,\psi} \; . \label{III}
\end{align}
In these three equations we used integration by parts (Wick's Theorem), and we manipulated the expressions in order to let the order parameters $q_{12}$ and $p_{12}$  appear (for their general definitions see Eqs.~\eqref{overlapneuroni} and~\eqref{overlapz}).
Term $IV$ is easily computed through the standard Gaussian integration:
\begin{align} \label{IV}
\begin{split}
 IV & = \frac{1}{N} \E \Biggl[ \frac{1}{Z_{N,k,p}(t,x,\psi)} \sum_\sigma \int d\MG(z)\ \frac{-\mathcal{C}}{2} \summup \zmu^2 B_{N,k,p}(t,x,\psi) \Biggr] = \\
    & = \frac{-\mathcal{C}}{2N} \summup \E \omega_{t,x,\psi} (\zmu^2) \; .
\end{split}
\end{align}

Summing the final expressions of Eqs.~\eqref{I},~\eqref{II},~\eqref{III} and~\eqref{IV} we have:

\begin{equation*} \label{dertot}
 \begin{split}
  \frac{d A_{N,k,p}}{d \psi} (t,x,\psi) = &  \frac{1}{2N} \bigl( \beta - \mathcal{B}^2 - \mathcal{C} \bigr) \summup \E \omega_{t,x,\psi} (\zmu^2) + \\
            & - \frac{\alphaN\beta}{\lb q_{12} p_{12} \rb_{t,x,\psi}} -\frac{\mathcal{A}^2}{2} \bigl( 1 - \lb q_{12} \rb_{t,x,\psi} \bigr)
              + \frac{\alphaN \mathcal{B}^2}{2} \lb p_{12} \rb_{t,x,\psi} \; ,
 \end{split}
\end{equation*}

which is what we reported in Eq.~\eqref{ss}.

\end{document}